\documentclass[]{aastex}
\usepackage{emulateapj5}
\usepackage{epsfig}
\newcommand{\LCDM}{$\Lambda$CDM}
\newcommand{\WMAP}{{\it WMAP}}
\newcommand{\GOODS}{{\it GOODS}}

\begin{document}
%\slugcomment{{\em submitted to Astrophysical Journal Letters}}

\shorttitle{The Redshift Distribution of Near-IR Selected Galaxies in \GOODS}

\shortauthors{Somerville et al.} 

\title{The Redshift Distribution of Near-IR Selected Galaxies in
the {\it Great Observatories Origins Deep Survey}
as a Test of Galaxy Formation Scenarios \altaffilmark{1}}

%-------------------------------------------------------------
\author{
Rachel S. Somerville\altaffilmark{2}, 
Leonidas A. Moustakas\altaffilmark{2}, 
Bahram Mobasher\altaffilmark{2}, 
Jonathan P. Gardner\altaffilmark{3}, 
Andrea Cimatti\altaffilmark{4}, 
Christopher Conselice\altaffilmark{5}, 
Emanuele Daddi\altaffilmark{6}, 
Tomas Dahlen\altaffilmark{2}, 
Mark Dickinson\altaffilmark{2}, 
Peter Eisenhardt\altaffilmark{7}, 
Jennifer Lotz\altaffilmark{8}, 
Casey Papovich\altaffilmark{9},
Alvio Renzini\altaffilmark{6},
Daniel Stern\altaffilmark{7} 
}

\altaffiltext{1}{Based on
observations taken with the NASA/ESA Hubble Space Telescope, which is
operated by the Association of Universities for Research in Astronomy,
Inc.\ (AURA) under NASA contract NAS5--26555, and on observations
collected at the European Southern Observatory, Chile, Programs
164.O-0561, 169.A-0725, 267.A-5729}
\altaffiltext{2}{Space Telescope Science Institute, 3700 San Martin
  Drive, Baltimore MD 21218; somerville,leonidas,mobasher,dahlen,med@stsci.edu}
\altaffiltext{3}{Laboratory for Astronomy and Solar Physics, 
Code 681, Goddard Space Flight Center, Greenbelt MD 20771; jonathan.p.gardner@nasa.gov}
\altaffiltext{4}{Osservatorio Astrofisico di Arcetri, I-50125 Firenze,
Italy; cimatti@arcetri.astro.it}
\altaffiltext{5}{California Institute of Technology, Mail Code 105-24, 
Pasadena CA 91125; cc@astro.caltech.edu}
\altaffiltext{6}{European Southern Observatory, D-85748, Garching, Germany; edaddi,arenzini@eso.org}
\altaffiltext{7}{Jet Propulsion Laboratory, California Institute of
  Technology, Mail Stop 169-506, Pasadena, CA 91109; prme@kromos.jpl.nasa.gov; stern@zwolfkinder.jpl.nasa.gov}
\altaffiltext{8}{Department of Astronomy, University of California, 
Santa Cruz, CA 95064; lotz@stsci.edu}
\altaffiltext{9}{Steward Observatory, University of Arizona, Tucson, AZ 85721; papovich@as.arizona.edu}

\begin{abstract}
The redshift distribution of near-IR selected galaxies is often used
to attempt to discriminate between the classical view of galaxy
formation, in which present-day luminous galaxies were assembled at
early times and evolve due to the passive aging of their stellar
populations, and that of hierarchical structure formation, in which
galaxies were assembled more recently via the merging of smaller
objects. We carry out such a test here, by computing the distribution
of photometric redshifts of $K_{\rm AB}<22$ galaxies in the Great
Observatories Origins Deep Field Survey (\GOODS) Southern Field, and
comparing the results with predictions from a semi-analytic model
based on hierarchical structure formation, and a classical `passive
evolution' model. We find that the redshift distributions at $z\la1.5$
of both the hierarchical and passive models are very similar to the
observed one. At $z\ga1.5$, the hierarchical model shows a deficit of
galaxies, while the passive model predicts an excess. We investigate
the nature of the observed galaxies in the redshift range where the
models diverge, and find that the majority have highly disturbed
morphologies, suggesting that they may be merger-induced
starbursts. While the hierarchical model used here does not produce
these objects in great enough numbers, the appearance of this
population is clearly in better qualitative agreement with the
hierarchical picture than with the classical passive evolution
scenario. We conclude that the observations support the general
framework of hierarchical formation, but suggest the need for new or
modified physics in the models.

\end{abstract}
\keywords{galaxies: formation --- galaxies: evolution}

%\subjectheadings{galaxies: formation --- galaxies: evolution}

%=======================
% 1
\section{Introduction}
\label{sec:intro}
%=======================

While the stars that produce the bulk of the optical light in nearby
luminous galaxies are known observationally to be quite old, it is
possible that the time at which the mass was assembled into a single
object is different from the formation time of the stars. In fact, in
the hierarchical paradigm of galaxy formation, one expects mass
assembly to be a gradual process, in contrast to the classical
monolithic dissipative collapse picture \citep*{els}. A generic
prediction of hierarchical (Cold Dark Matter; CDM) theories is that
galaxies should be less massive in the past. Direct searches for
galaxies at high redshift should therefore provide a crucial test of
this class of theories \citep{kc98}. Determining the epoch of
formation and assembly of present-day luminous galaxies is central to
our understanding of galaxy formation and cosmology, and is a primary
goal of the Great Observatories Origins Deep Survey (\GOODS), as well
as of many other deep surveys.

In practice, however, there are many complications involved in
carrying out this test. Of course, we cannot follow an individual
galaxy back in time, but can only study populations observed at
different redshifts.  Cosmological k-corrections (redshifting of light
to longer wavelengths) and stellar evolution both make it difficult to
relate high redshift populations to local ones. One can attempt to
account for these effects by artificially redshifting template spectra
for representative galaxy types and/or by using stellar population
models with assumed star formation histories. Starting from an
empirical low-redshift luminosity function, one can then predict how
the local galaxy population would appear at high redshift, with the
effects of k-corrections only (`no evolution') or with the additional
effects of stellar evolution (`passive evolution') included. However,
these corrections can be sensitive to the input assumptions.

There are several advantages to carrying out this test using a sample
selected in the near-IR rather than the optical. The corrections
described above are considerably smaller, and the near-IR light more
closely traces the stellar mass. The observed K-band (2.2 $\mu$m)
probes the SED longwards of the rest-frame I-band ($\sim 8600$ \AA)
out to $z\sim 1.5$. However, it also necessary to probe a large enough
volume to contain a statistically significant sample of rare luminous
objects. The availability of accurate photometric redshift estimates,
requiring multi-band U through K photometry, enables such tests to be
extended into the $z\ga 1.2$ `spectroscopic desert', where
spectroscopic redshifts are difficult to procure. Only recently have
sufficiently deep and wide near-IR selected surveys with multi-band
photometry begun to become available.

The original study by \citet[][KC98]{kc98} showed that the cumulative
redshift distribution of K-selected galaxies differed greatly in a
hierarchical model compared with a passive (or Pure Luminosity
Evolution; PLE) model: they found that the fraction of galaxies at
$z>1$ with $18 < K_{\rm VEGA} < 19$ was an order of magnitude higher
in the PLE model. Based on the small observational samples available
at the time, KC98 concluded that the hierarchical model provided a
better match to the observed redshift distribution out to
$z\sim1$. Recently, several studies have carried out this test using
updated hierarchical models (including, among other things, the
transition to a low-$\Omega_m$, cosmological constant-dominated
cosmology) and larger, deeper observational samples
\citep{firth,cimatti:02b,kashikawa:03}. Both \citet{cimatti:02b} and
\citet{kashikawa:03} concluded that the PLE models produced better
agreement with the observed redshift distribution at $z\ga1.5$ than
the hierarchical models, although the disagreement was relatively
subtle compared with the expectations set out in KC98.

In this paper, we use a K$_s$-band selected sample ($K_{AB}<22$) from
the \GOODS\ Southern field to repeat the KC98-type redshift
distribution test, using accurate, well-calibrated photometric
redshifts \citep{mobasher:03}. The \GOODS\ field probes a
considerably larger area and volume than previous studies at a similar
depth. We also have the advantage of the exquisite ACS imaging, which
allows us to investigate the morphologies of high redshift galaxies,
gaining further insights into the nature of these objects. We confront
these observations with predictions from semi-analytic hierarchical
galaxy formation models, and with PLE models, both normalized to the
$z=0$ K-band luminosity function recently determined from the 2MASS
survey \citep{kochanek:01,cole:01}.

\section{The Data}
The \GOODS\ data are described in \citet{giavalisco:03}.  Our study is
based on the \GOODS\ Southern field (Chandra Deep Field South; CDFS),
which has an area of 160 arcmin$^2$. In addition to the 4-band (BViz)
ACS imaging, we make use of an extensive set of complementary
ground-based observations from the VLT, NTT, and ESO 2.2m telescopes,
including optical WFI (U'UBVRI) and FORS (RI), and infrared SOFI
(JHK$_{s}$) photometry, which covers the entire ACS \GOODS\ CDFS
field. This work is based on a PSF-matched SOFI-K$_s$ selected
catalog.  Clearly-unresolved sources (stars) based on the ACS
$z_{850}$ data down to z$_{AB}<26.2$ have been removed.  The 50\%
completeness limit is K$_{s}=22.8$, and the sample should be close to
100\% complete at K$_s<22$ \citep{moy:03,giavalisco:03}, the limit we
adopt for our analysis. Photometric redshifts were estimated as
described in \citet{mobasher:03}, using all the available bands (U'
through K$_s$), and are well-calibrated to $K_{\rm VEGA}<20$ using
spectroscopic redshifts from the K20 survey \citep{cimatti:02} and
additional spectra obtained from FORS2 on the VLT as part of the
\GOODS\ program. Typical redshift errors for the K$_s<22$ sample are
$\sim \Delta z/(1+z_{\rm spec}) = 0.1$ \citep{mobasher:03}.

In the remainder of the paper, we refer to the $K_s$ band as $K$ for
brevity, and give all magnitudes in the AB system unless otherwise
specified.  Note that for our filter bands, $K_{AB} = K_{VEGA}+1.85$,
and $(R-K)_{AB}= (R-K)_{VEGA}-1.65$.

\section{Models}
Where relevant, we assume the following values for the cosmological
parameters: matter density $\Omega_m = 0.3$, baryon density $\Omega_b
=0.044$, dark energy $\Omega_{\Lambda}=0.70$, Hubble parameter
$H_0=70$\,km\,s$^{-1}$\,Mpc$^{-1}$, fluctuation amplitude $\sigma_8 =
0.9$, and a scale-free primordial power spectrum $n_s=1$. These values
are consistent with the recent \WMAP\ data \citep{spergel:03}.

The basic ingredients of the semi-analytic hierarchical models used
here are described in \citet{sp} and \citet{spf}. The models are based
on hierarchical merger trees within a \LCDM\ model, and include
modeling of gas cooling, star formation, supernova feedback, chemical
enrichment, stellar population synthesis, and dust.  We use the
multi-metallicity stellar SED models of \citet{dgs}, assuming a
Kennicutt IMF. Here we have considered a model based on the
`collisional starburst' recipe described in \citet{spf}, which was
found to produce the best agreement with high redshift ($z\sim3$)
galaxy observations. Several parameters and model ingredients have
been adjusted to give better agreement with the low redshift optical
and K-band luminosity functions recently determined by SDSS and 2MASS
\citep[e.g.][]{blanton:01,cole:01}, and with low redshift galaxy
colors (details will be given in Somerville et al. 2003, in prep). We
produced a mock catalog with the same angular extent and depth as the
\GOODS\ ACS and ground-based data, which was run `blind' before the data
were analyzed.

The passive evolution models are computed as described in
\citet{gardner:98}, and are normalized to the type-dependent $z=0$
K-band luminosity functions derived from the 2MASS survey by
\citet{kochanek:01}. These models contain six different types of
galaxies with simple parameterized star formation histories: E, S0,
Sb, Sc, Irr, and starburst. All galaxies except the starburst type
begin forming stars at $z=15$. The E, S0, Sb and Sc types have
exponentially declining star formation rates with e-folding timescales
of 1 Gyr for the E/S0, 4 Gyr for the Sb and 7 Gyr for the Sc
types. The Irr types have a constant star formation rate. The
starburst population has a constant star formation rate and constant
age (1 Gyr) at every redshift.

\section{Results}

%\clearpage

\begin{figure*} 
\epsscale{2.0}
\plotone{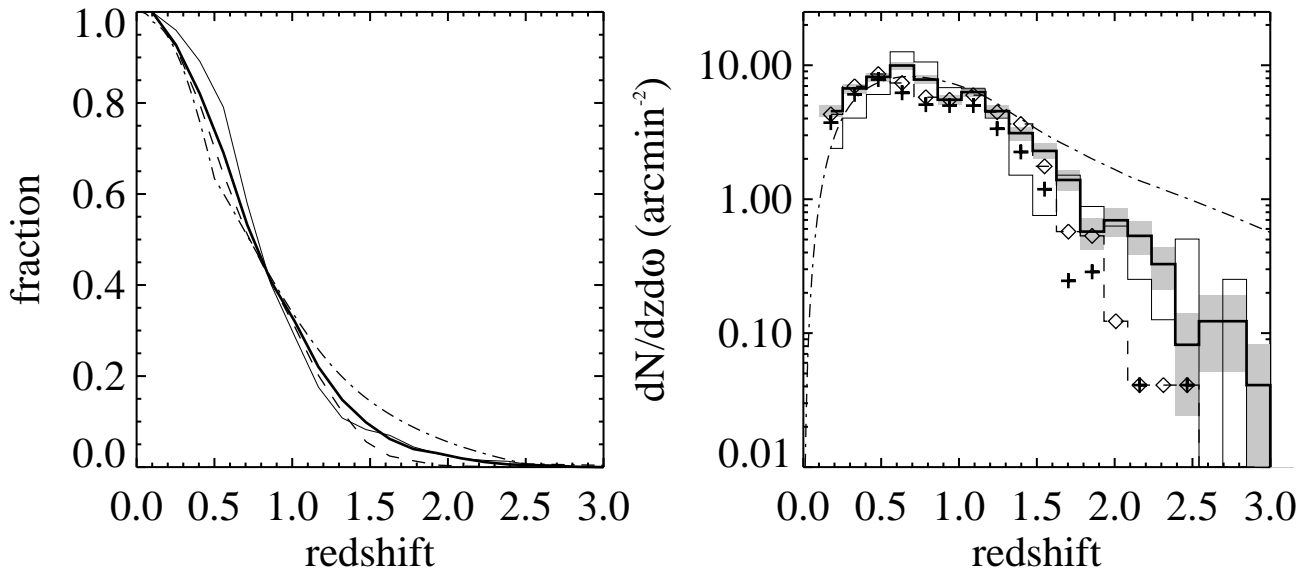}
\caption{\small [left] Cumulative redshift distribution of $K<22$
galaxies. Bold solid line: \GOODS\ CDFS; Light solid line: K20 sample;
dashed: semi-analytic model; dot-dashed: passive model. [right]
Differential redshift distribution ($K<22$), with Poisson errors shown
on the \GOODS\ data [bold histogram]. Line types are as in the left
panel. The hierarchical model is highlighted by diamond symbols. The
cross symbols show the hierarchical model with a crude correction of
0.2 magnitudes applied to account for the isophotal magnitudes used in
the \GOODS\ observations (see text).
\label{fig:nofz}}
\end{figure*}

%\clearpage

The cumulative and differential redshift distributions for the $K<22$
selected samples are shown in Fig.~\ref{fig:nofz}, for the \GOODS\
data, the K20 survey, and the semi-analytic and passive models.  The
redshift distribution of \GOODS\ agrees very well with that obtained
from the K20 spectroscopic survey, which was carried out independently
in a smaller area of the same field. The agreement between the
semi-analytic models and the observations is quite good --- well
within the fluctuations expected from large scale structure --- up to
about $z\sim1.5$. This good low redshift agreement is in contrast with
previous comparisons with semi-analytic models from several groups
\citep{firth,cimatti:02b}. In previous models, the luminosity function
was too steep on the faint end, leading to an excess of intrinsically
faint galaxies at low redshift. Improved modeling of sub-halo merging,
and including ejection of gas by superwinds and suppression of gas
infall in small halos after reionization produce better agreement with
the observed K-band luminosity function at $z=0$
\citep{squelch,benson:02}, and also to better agreement with the
low-redshift $N(z)$ shown here. Also of note is the similarity of the
predicted $N(z)$ at $z<1$ for the hierarchical and the PLE models ---
in contrast with the results of KC98.

At $z\ga 1.5$, the number of $K<22$ galaxies in the semi-analytic
models is significantly and systematically smaller than the observed
value, while the PLE model systematically \emph{overpredicts} the
number of objects in this range by a similar factor. The isophotal
magnitudes used in our \GOODS\ catalog are probably fainter than the
true total magnitudes by about 0.2--0.3 magnitudes
\citep{cimatti:02}. Correcting the semi-analytic models for this
effect further exacerbates the discrepancy, as shown in
Fig.~\ref{fig:nofz}. We note that the redshift range in which the
models suffer from the most significant discrepancy ($z\ga 1.5$) is
precisely where the photometric redshifts are the least secure, as
very few spectroscopic redshifts are available to test them --- though
progress is being made in this area \citep{daddi:03}. In addition, at
these redshifts, the observed K-band has shifted into the rest
optical, and is therefore more sensitive to recent star formation
activity.

\begin{figure*} 
\epsscale{2.0}
\begin{center}
\plottwo{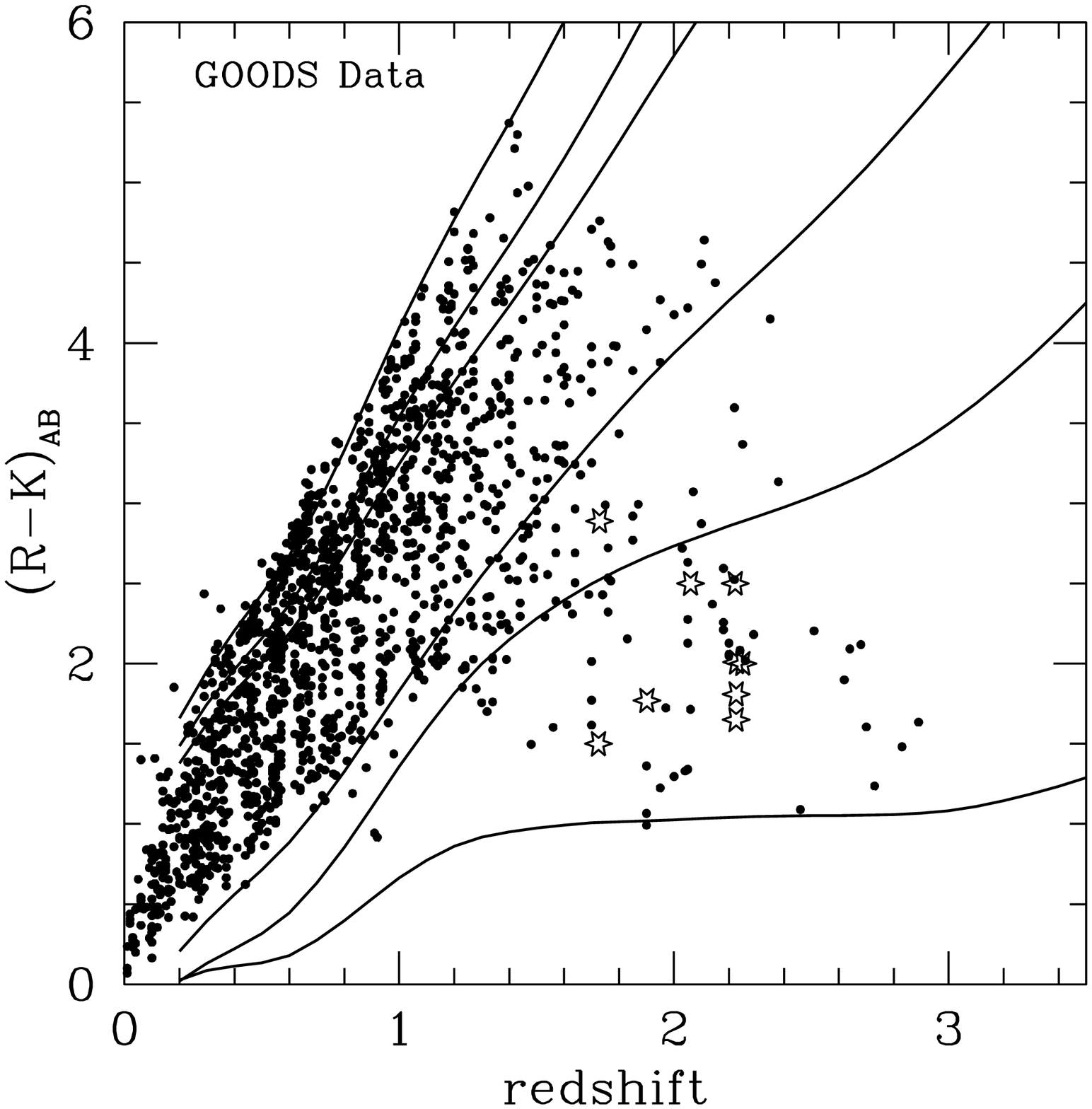}{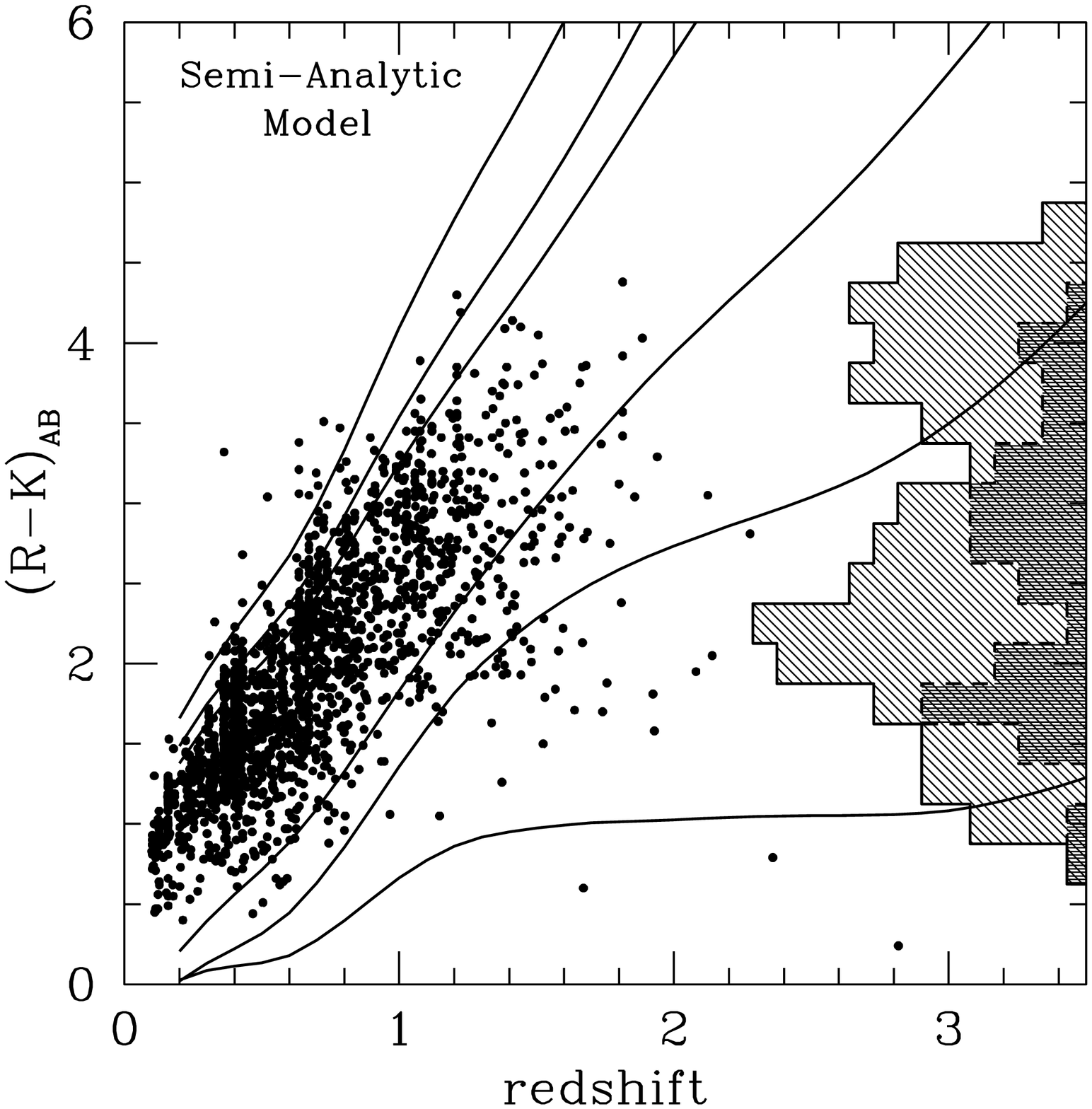}
\end{center}
\figcaption{\small Observed frame color $(R-K)_{AB}$ vs redshift for
the \GOODS\ CDFS [left] and the semi-analytic mock catalog
[right]. Tracks are shown for single age, stellar metallicity
populations with ages of 13.5, 5.8, 3.2, 1, 0.5, and 0.1 Gyr
(unreddened), from top to bottom. Large star symbols show the colors
of spectroscopically confirmed objects from
\protect\citet{daddi:03}. The color distribution of galaxies with
$K<22$ in the redshift range $1.7 < z < 2.5$ is shown in the right
panel as histograms (diagonal hatched: \GOODS; horizontal fill: mock
catalog).
\label{fig:rminusk}}
\end{figure*}

It is interesting to investigate the nature of the galaxies that
appear in the high redshift tail of the observed distribution, but are
under-represented in the hierarchical models. Fig.~\ref{fig:rminusk}
shows the (observed frame) $R-K$ colors as a function of redshift for
the observed and semi-analytic model galaxies. The colors of solar
metallicity, single-age populations are also shown. The semi-analytic
model reproduces the locus of observed colors fairly well up to
$z\sim0.8$. At $z\sim0.8$--1.2, the semi-analytic model produces
enough $K<22$ galaxies overall, but does not produce enough galaxies
with very red colors. This problem has been noted before
\citep{dcr:00,firth,cimatti:02b}. The number densities and
morphologies of these $R-K \ga 3.35$ Extremely Red Objects (EROs) in
\GOODS\ are discussed in more detail in \citet{moustakas:03}.

At higher redshift ($z\ga 1.5$), the observed distribution of $R-K$
colors is bimodal, with the `dip' around $R-K\sim3$ (see
Fig.~\ref{fig:rminusk}). Focusing on galaxies in the redshift interval
$1.7 < z < 2.5$ and with $K<22$, we find that about 60\% (44 out of
75) of the \GOODS\ galaxies have colors bluer than $R-K=3$, while in
the semi-analytic mock catalog, 52\% (11 out of 21) of the objects
have $R-K<3$. Considering the small number statistics, this implies
that the semi-analytic model produces approximately the correct
\emph{relative fraction} of red and blue galaxies.
%; or put another way,
%that the observed high redshift galaxies causing the discrepancy are
%nearly evenly divided between red and blue objects. 
We have visually inspected all of the objects in this sub-sample, and
find that the great majority of both red and blue galaxies have highly
irregular morphologies, many with multiple components and the
appearance of ongoing mergers.

From comparison with the single-burst model tracks, we can deduce that
the high redshift, blue galaxies ($R-K \la 3$) must be dominated by
extremely young ($\la 500$ Myr), nearly unreddened stars.
% (or, that these galaxies have mis-assigned photometric redshifts).
Intriguingly, \citet{daddi:03} have recently obtained spectra for a
sample of galaxies with $K_{\rm VEGA}<20$ and $z_{\rm phot} > 1.7$ in
the \GOODS\ CDFS/K20 field, and have successfully obtained redshifts
for 9 such objects, confirming that they lie in the range $1.7 \la z
\la 2.3$. On the basis of these spectra and the ACS images,
\citet{daddi:03} argue that these objects are strongly clustered,
massive, merger-driven starbursts. We show the location of these
objects on our color-redshift diagram, and see that they lie precisely
in the regime of the `missing blue galaxies'.

%\clearpage

%\clearpage

\section{Discussion}

In this paper, we address a central question in galaxy formation
theory: were most of the luminous galaxies that we see today already
in place at high redshift, or were they assembled gradually over time?
To answer this question, we confronted observations from the \GOODS\
CDFS field with predictions from two models representing what are
traditionally considered opposing points of view: a semi-analytic
hierarchical model based on CDM theory, and a `passive evolution'
model, in which galaxy properties evolve only due to the aging of
their stellar populations. Our main conclusions are as follows:

\begin{enumerate}
\item Up to $z\sim1$, redshift distributions of $K<22$ galaxies in the
hierarchical model, the passive model, and the data are all consistent
with one another. However, the hierarchical model underproduces the
number of Extremely Red Objects (EROs) at $z\sim1$.

\item The hierarchical model underproduces near-IR selected objects
($K<22$) by about a factor of three at $z\ga1.7$ and by an order of
magnitude at $z\ga 2$. The PLE model overproduces these galaxies by
about a factor of two at $z\sim2$.

\item At $z\ga 1.5$, the objects underproduced in the hierarchical
model are nearly equally divided between red ($R-K>3$) and blue
galaxies. Based on ACS imaging, many of these objects appear to be
highly morphologically disturbed, and a large fraction may be
merger-driven starbursts.

\end{enumerate}

Not surprisingly, the predicted colors of model galaxies in the
hierarchical models are quite sensitive to the details of the star
formation recipes, as well as the stellar IMF and dust modeling. For
example, if we assume that starbursts occur in major mergers only, we
can produce more extremely red galaxies at $z\sim1$, but we then
produce even fewer luminous blue galaxies at
$z\sim1.5$--2. Alternatively, if we brighten all model galaxies by 0.5
magnitudes (40\%), we find that the semi-analytic model then produces
sufficient numbers of $K<22$ objects at $z\ga1.5$, but this naturally
causes an excess at lower redshift.  Cosmic variance is also expected
to be significant for these luminous, rare objects --- assuming that
these objects are strongly clustered, like EROs at $z\sim 1$
\citep[e.g.][]{daddi:01}, we estimate an uncertainty due to cosmic
variance of about 60\% in the number density of objects at $1.5 \la z
\la 2$ \citep[see][]{cosvar}. This implies that the semi-analytic
model is discrepant at less than 2$\sigma$. Results from additional
fields will determine whether there is an overdensity of objects at
$z\ga1.5$--2 in the CDFS.

A significant conclusion from this work is that the test proposed by
KC98, when carried out with recent models, is not as strong a
discriminator between the traditionally opposing points of view of
hierarchical vs. PLE models as was found in that work. Adoption of the
flat, low-$\Omega_m$ cosmology now favored by observation, and the
refinement of the star formation and feedback recipes has resulted in
more early star formation in the modern semi-analytic models. At the
same time, the use of the observed K-band $z=0$ luminosity function to
normalize the PLE models has reduced the uncertainty due to dust and
k-corrections in those models. The net effect is that the two
scenarios diverge significantly only at a higher redshift ($z\ga1.5$)
than predicted by KC98. Several other recent studies
\citep{cimatti:02,firth,kashikawa:03} have reached a similar
conclusion.

However, the morphologies of the observed objects in this redshift
interval are inconsistent with the passive evolution hypothesis ---
the majority seem to be highly disturbed morphologically, and many are
clearly interacting or merging \citep[see
also][]{daddi:03}. Qualitatively, this is clearly more consistent with
the hierarchical scenario. However, the \emph{quantitative}
disagreement between the number of predicted and observed objects
indicates that some ingredients in the models need to be modified, or
else that some physics is missing. Further study of the nature of this
population at $1.5 \la z \la 2.2$, which forms a `bridge' between the
better-studied populations of `normal' galaxies at $z\la1$ and
Lyman-break galaxies at $z\ga 2.2$, will certainly provide important
new insights into some of the remaining mysteries of galaxy formation.

%===================================
\section*{Acknowledgments}
\begin{small}

We thank our collaborators in the \GOODS\ team for useful feedback on
this work. Support for this work was provided by NASA through grants
GO09583.01-96A and GO09481.01-A from the Space Telescope Science
Institute, which is operated by the Association of Universities for
Research in Astronomy, under NASA contract NAS5-26555. Support for
this work, part of the {\it Space Infrared Telescope Facility (SIRTF)}
Legacy Science Program, was also provided by NASA through Contract
Number 1224666 issued by the Jet Propulsion Laboratory, California
Institute of Technology under NASA contract 1407.

\end{small}
%=====================================

\bibliographystyle{apj} 
\bibliography{apj-jour,massgal}

\end{document}